\documentclass[12pt,journal]{IEEEtran}
\usepackage{setspace}

\usepackage{multirow}
\usepackage{epsfig}
\usepackage{rotating}
\usepackage{subfigure}
\usepackage{color}
\usepackage{algorithmic}
\usepackage{algorithm}
\usepackage{lscape}
\usepackage{spk}
\newtheorem{definition}{Definition}

\begin{document}
\title{Inference of Gene Predictor Set Using Boolean Satisfiability}
\author{
Pey-Chang Kent Lin \hspace{0.8in} Sunil P Khatri\\
k1arte@neo.tamu.edu \hspace{0.4in} sunilkhatri@tamu.edu\\
Department of Electrical \& Computer Engineering \\Texas A\&M University,
College Station TX 77843\\
}

\date{}
\maketitle
\thispagestyle{empty}



\begin{abstract}
The inference of gene predictors in the gene regulatory network
has become an important research area in the genomics and medical 
disciplines. Accurate predicators are necessary for constructing
the GRN model and to enable
targeted biological experiments
that attempt to confirm or control the regulation process. In this
paper, we implement a SAT-based algorithm to determine the gene 
predictor set from steady state
gene expression data (attractor states). 
Using the attractor states as input, the states
are ordered into attractor cycles. For each attractor cycle ordering,
all possible predictors
are enumerated and a CNF expression is formulated which encodes these
predictors and their
biological constraints. Each CNF is explored using a
SAT solver to find candidate predictor sets. Statistical analysis 
of the results selects the most likely predictor set of the GRN
corresponding to the attractor data. We demonstrate our algorithm 
on attractor state data from a melanoma study~\cite{bittner00} and present our
predictor set results.
\end{abstract}


\section{Introduction}
\label{sec:intro}

With the mapping of the human genome complete, the focus in computational 
biology has shifted from sequence analysis to the
understanding of 
gene regulation and its inter-relation with the biological system.
The use of genome information has given rise to 
the notion of "personalized medicine" --
targeted and specific disease prevention and treatment based on individual
gene information~\cite{burke07, teutsch09}.
The urgent applications to cancer and gene-related diseases calls for
the genomics field to significantly improve the algorithms used for 
accurate inference of the gene regulatory network (GRN). 

In an organism, the genome is a highly complex control system wherein 
proteins and RNA produced by genes interact with and regulate the activity 
of other genes~\cite{guelzim02}. 
The activity of a target gene $g_i$
is regulated (or predicted) 
by the genes in its {\em predictor} (e.g. if $g_1$ becomes
inactive when $g_2$ and $g_3$ are active, then $g_2$ and $g_3$ are
called predictors of g). 
The complete set of predictors 
({\em predictor set}), which contains the predictors for each gene in 
the GRN, describes the interaction of all genes
within the gene regulatory network and is the prerequisite for inferring 
the GRN structure.

There are several GRN characteristics that impact the formulation of our
GRN model and predictor inference algorithm.
First, the gene activity level of all genes at a 
particular time $t$ represents the {\em state} of the GRN at that time $t$.
From our knowledge of biological systems, we 
observe
that over time, cellular processes
transition to stable {\em attractor} states. 
Some of these attractor states represent normal cellular phenomena
in biology such as cell cycle and division. 
However, some attractor
states are consistent with disease such as the metastasis of cancer.
Second, the GRN is often inferred by observing
microarray-based experimental data which measures the activity level of 
genes. The correlation of the observed gene activity (or state) can be
used to help describe the gene regulation.
The disadvantage of using microarray data is such that studies do not involve
controlled time experimental data (time-series data). Hence the measurements 
are
assumed to arise from the cyclic sequence of gene expressions
(attractor states) in steady state
({\em attractor cycles}). 
The GRN is then inferred from this data, using methods 
traditionally based on probabilistic transition models~\cite{dougherty00, zhao08}.

As previously mentioned, it is necessary to determine the predictor set to
reconstruct the GRN. However, there may exist many possible predictors
for any gene,
based on the attractor cycle data. Furthermore, only certain
combinations of predictors may form a valid predictor set due to 
biological constraints. The issue addressed in this paper is
 how to efficiently and deterministically
select the predictors that form the predictor set.
We have implemented a Boolean satisfiability (SAT) 
based algorithm for the inference of gene predictors. Satisfiability is a 
decision problem of determining whether the variables in a Boolean 
formula 
(expressed in Conjunctive Normal Form or CNF)
 can be assigned to make the formula evaluate to {\em true}. 
Although SAT is NP-complete, many SAT solvers have been developed 
to quickly and efficiently solve large SAT 
problems. Our algorithm takes advantage of advanced SAT solvers to find 
the predictor set.

The basic outline of our SAT-based algorithm is described 
briefly below.
First, all possible
orderings of attractor state are
enumerated, yielding all possible attractor cycles. For each ordering, we enumerate
all predictors that are logically valid, and create a CNF expression 
which encodes all these predictors and biological constraints
(such as cardinality bounds on the predictors). A SAT solver 
is used to find the valid candidate
predictor sets. After this process is done 
iteratively for all attractor cycle (orderings), statistical analysis provides 
the most likely candidates for the predictor set.

The key contributions of this paper are:
\begin{itemize}
\item We develop a 
Boolean Satisfiability based approach to realize 
the gene predictor set from attractor state data.
\item We modify an existing SAT-solver (MiniSat~\cite{minisat}) 
for efficient all-SAT computation and
further optimize MiniSat for improved predictor inference.
\item On gene expression data from 
a melanoma study~\cite{bittner00}, we apply our SAT-based algorithm and 
present results for genes that regulate all the genes, including
the cancer gene WNT5a.
\item Our approach can be used to find the predictor set for any gene 
related disease, provided attractor state data is available. The 
predictor set information
obtained from our algorithm can be used by biologists to fine tune their
gene expression experiments.
\end{itemize}

The remainder of this paper is organized as follows. 
Section~\ref{sec:previous_work} describes previous work in modeling the 
gene regulatory network and inference of gene predictors. 
Section~\ref{sec:our_approach} presents 
our FSM model and Boolean SAT approach.
Section~\ref{sec:expt} reports experimental results. 
Concluding comments and future work are discussed in Section~\ref{sec:conclusions}.

\section{Previous Work}
\label{sec:previous_work}
Several models have been proposed for modeling the GRN such as 
Markov Chains~\cite{kim02, vahedi08},
Coupled ODEs (ordinary differential equations), 
Boolean Networks~\cite{kauffman69, shmulevich09},
Continuous Networks~\cite{geard05}, and Stochastic Gene Networks~\cite{arkin98}. 

This paper utilizes the Boolean Network (BN) model that was proposed
by Kauffman in 1969~\cite{kauffman69}. In a Boolean Network, the expression 
activity
of a gene is represented as a binary value, where 1 indicates the
gene is ON (active) and producing gene-products, while 0 indicates it is 
OFF. Such a model cannot capture the continuous and stochastic biochemical
properties of protein and RNA production. However, genes can typically be
modeled as ON or OFF in any particular biochemical pathway. 

In~\cite{zhou04, zhou06}, the probabilistic modeling framework is 
represented by
dynamic Bayesian networks and probabilistic Boolean networks (PBNs).
The method proposed considers gene prediction
using multinomial probit regression with Bayesian variable selection.
Genes are selected which satisfy multiple regression equations, of which the
strongest genes are used to construct the predictor set. The
target gene is predicted based on the strongest genes, using
the coefficient of determination to measure predictor accuracy. 

Another method proposed by~\cite{pal05} also assumes PBN.
A partial state transition table is constructed based on available
attractor state data. From this state transition table, predictors with 3 or
less regulating
genes are selected for each target gene. All unknown values in
the table are randomly set. The Boolean network is simulated for several iterations 
on several starting states, observing whether the states eventually transition
to an attractor cycle. 
If the simulation successfully transitions to attractor cycles, the
selected predictors are considered as a valid predictor set.
This process is repeated to build a collection of Boolean Networks which
are combined to form a Probabilistic Boolean Network (PBN).

Our larger goal is to
find a small number of deterministic 
GRNs, rather than a PBN.
Towards this, we need to find ways to accurately find the predictor set.
This is the focus of this paper. Philosophically, our aim is to invest 
effort into accurate predictor set determination, so that the results can be
used to find high quality deterministic GRNs.

\section{Our Approach}
\label{sec:our_approach}

This section describes our model and algorithm for inference of predictor
sets
using SAT. We begin with some logic synthesis definitions which are useful in
understanding the application of SAT to GRNs
 and predictor selection. We then
describe a simple example to explain the algorithm.
Lastly, we generalize the algorithm for
larger problem sets and comment on specific issues 
about the use of SAT and complexity.

\subsection{Definitions}

\begin{definition}
A {\bf literal} or a {\bf literal function} is a binary variable $x$ or its 
negation $\overline{x}$. 
\end{definition}

\begin{definition}
A {\bf cube} is a product of a set of literal functions.
\end{definition}

\begin{definition}
A {\bf clause} is a disjunction containing literals.
\end{definition}

\begin{definition} A {\bf Conjunctive Normal Form (CNF)} expression
consists of a conjuction (AND) of $m$ clauses $c_1 \ldots c_m$. Each
clause $c_i$ consists of disjunction (OR) of $k$ number of literals. 
\end{definition}

A CNF formula is also referred to as a logical product of sums. Thus, to satisfy the
formula, each clause must have at least one literal evaluate to true.

\begin{definition}
{\bf Boolean satisfiability (SAT)}.
Given a Boolean formula $S$
on a set of binary variables $X$
, expressed in Conjunctive Normal Form (CNF),
the objective of SAT is to
identify an assignment of the binary variables in X
that satisfies $S$, if such an assignment exists.
\end{definition}
For example, consider the formula $S(a,b,c) = 
(a+\overline{b})\cdot(a+b+c)$. This formula consists of 3 variables, 2 clauses,
and 4 literals. This particular formula is satisfiable, and a satisfying 
assignment is $<a,b,c>\,=\,<0,0,1>$, which can be expressed as
the satisfying cube $\overline{a}\overline{b}c$.

There may exist many satisfying assignments for
the formula in question. An 
extension of the SAT problem is to find {\em all} satisfying assignments (or
All-SAT). One simple method to accomplish All-SAT is to repeatedly 
run SAT on the formula $S$, express each satisfying assignment as a
cube $k$, complement $k$ to get a clause $c$, and add $c$ as a new clause of the
formula
and running SAT again. 
The inclusion of $c$ in $S$ ensures that the same cube $k$ cannot
be found as a satisfying assignment again.
The process continues until no new solutions can be found. In the 
previous example, the satisfying 
cube $\overline{a}\overline{b}c$ is complemented
and added as a new clause $(a+b+\overline{c})$ to the
original formula to be solved by SAT again (this is repeated until no new
satisfying assignments are found).

\begin{definition}
A {\bf predictor}
$f_i=\{g_j,g_k,\cdots\}$
lists the set $\{g_j,g_k,\cdots\}$ of genes which regulate the activity
of gene $g_i$.
\end{definition}

\begin{definition}
The {\bf predictor set} is the complete set of predictors 
\{$f_1, f_2, \cdots, f_n$\}
for the GRN with $n$ genes $g_1,g_2,\cdots,g_n$.
\end{definition}

\subsection{Implementation and Example}
Given gene expression data (a set of attractor states) as input,
we would like to determine the best predictor set.
We first present an outline of our SAT-based algorithm,
and then explain the steps through a simple example.

The algorithm has three main steps.
\begin{itemize}
\item First, attractor states are ordered into
attractor cycles. For each possible ordering
of the attractor states in to attractor cycles, all possible predictors are 
found and a CNF is generated containing the predictors and constraints. 

\item Second, the CNF is solved for All-SAT, recording all satisfying cubes. 
Each cube corresponds to a predictor set.
The first two steps are repeated for all attractor cycle orderings. 

\item Finally, statistical analysis
on the SAT results determines the most frequent (likely) predictor set
for the GRN.
\end{itemize}

We apply the SAT-based algorithm to a simple example with three
genes $(g_1, g_2, g_3)$ and gene expression data with two lines $(000, 101)$. 
The present state of these genes is represented by the variables 
$<x_1, x_2, x_3>$ and the next state is represented by the variables
$<y_1, y_2, y_3>$.
We assume each line was measured in steady state and therefore is an
attractor state.

{\bf Step 1:} We order (or arrange) the
attractor 
states into valid attractor cycles, of which there are two 
possibilities. One ordering is with each attractor
state transitioning to itself with a self-edge, thus 
resulting in an attractor cycle 
of length one, as shown in Table~\ref{tab:stt}. 
The other possible ordering is a transition from one state to the other
and back, forming a single attractor cycle of length two.

For each valid attractor cycle ordering, a 
{\em partial state transition table}
is constructed containing the attractor states. For example, if the first
attractor cycle ordering 
(in which states transition to themselves) is chosen, the resulting 
state transition table is shown in Table~\ref{tab:stt}. To find all valid
predictors of a gene, each next state column is checked against 
all combinations of the present state columns. For example with gene $g_1$,
the next state bit $y_1$ is $0$ in the first row and $1$ in the second
row. Hence the present state bit $x_2$ alone cannot predict $y_1$ as there
is a contradiction (since $y_1 = 0$ from the first row and
$y_1 = 1$ in the second row and $x_2 = 0$ for both rows). 
However if we consider state bits
$x_2$ and $x_3$,
we find that they together can predict $y_1$, 
since
the pair of values $<x_2,x_3>$ is different in the two rows.
Thus gene $g_1$ can be regulated by genes $g_2$ and $g_3$, so one valid
predictor for $g_1$ is $f_1 = \{x_2,x_3\}$.
All valid predictors with 3 or less inputs are exhaustively 
searched and recorded for CNF formulation
in the next step. In our example, gene $g_1$
has 5 possible predictors $\{x_3\}, \ \{x_1,x_2\}, \ \{x_1,x_3\},
\ \{x_2,x_3\}, \ \{x_1,x_2,x_3\}$ 
which we label $v^1_1, v^1_2, v^1_3, v^1_4$ respectively.
We assume that
a gene cannot self-regulate, so $\{x_1\}$ is not a valid predictor.

\begin{table}[tb]
\footnotesize
\begin{center}
\begin{tabular}{|c|c|c||c|c|c|}\hline
\multicolumn{3}{|c||}{\bf Present state} 
	& \multicolumn{3}{c|}{\bf Next state} \\ \hline
$x_1$ & $x_2$ & $x_3$
	& $f_1$ & $f_2$ & $f_3$ \\ \hline
0 & 0 & 0 
	& 0 & 0 & 0 \\ \hline
1 & 0 & 1
	& 1 & 0 & 1 \\ \hline
\end{tabular}
\end{center}
\caption{Example state transition table}
\label{tab:stt}
\end{table}

{\bf Step 2:} After all predictors are found, we generate
 the SAT formula
which encodes valid
predictor sets for all possible predictor combinations.
Each predictor is assigned a variable $v^i_j$ which
corresponds to the $j^{th}$ predictor for gene $i$. Gene $g_1$ in our example
will have five predictor variables $v^1_1 \equiv \{x_3\},
\, v^1_2 \equiv \{x_1,x_2\},
\, v^1_3 \equiv \{x_1,x_3\},
\, v^1_4 \equiv \{x_2,x_3\},
\, v^1_5 \equiv \{x_1,x_2,x_3\}$.
Gene $g_2$ will have the predictor variables
$v^2_1 \equiv \{x_1,x_2\},
\, v^2_2 \equiv \{x_1,x_3\},
\, v^2_3 \equiv \{x_2,x_3\},
\, v^2_4 \equiv \{x_1,x_2,x_3\}$.
Gene $g_3$ will have the predictor variables
$v^3_1 \equiv \{x_1\},
\, v^3_2 \equiv \{x_1,x_2\},
\, v^3_3 \equiv \{x_1,x_3\},
\, v^3_4 \equiv \{x_2,x_3\},
\, v^3_5 \equiv \{x_1,x_2,x_3\}$.
There are three constraints that we incorporate while 
constructing the CNF. The conjuction of these constraints our final CNF.

\begin{enumerate}
\item The first constraint $(S_1)$ is that all genes in the GRN
must have a predictor. In other words, we assume that all genes
are "participating" in the GRN and that all genes predict at least 
one other gene. 
For gene $i$, all of its associated predictor variables are
written in a single clause $c^1_i = (v^i_1 + \cdots + v^i_j)$. 
The clause for gene $g_1$ in our example is formulated
as $c^1_1 = (v^1_1 + v^1_2 + v^1_3 + v^1_4 + v^1_5)$.
To satisfy this satisfy this clause, at
least one predictor among $v^1_1, \cdots, v^1_5$ must be chosen. 
Then to ensure at least one predictor is chosen for all genes
we write the conjunction of all $c^1_i$ clauses.

$S_1 = c^1_1 \cdot c^1_2 \cdot c^1_3$

$S_1 = (v^1_1 + v^1_2 + v^1_3 + v^1_4 + v^1_5) \cdot 
(v^2_1 + v^2_2 + v^2_3 + v^2_4) 
\cdot (v^3_1 + v^3_2 + v^3_3 + v^3_4 + v^3_5)$

\item The second constraint $(S_2)$ specifies that for each gene, there 
exists only one predictor. A gene cannot be regulated by multiple
sets of predictors. To formulate the clauses $c^2_i$ for gene $i$, 
smaller clauses are formed from 
all pair combinations of its predictors $v^i_{1 \cdots j}$. In each of these
clauses of pair of variables, both predictor variables are complemented.
For gene $g_1$, $c^2_1 = 
(\overline{v^1_1} + \overline{v^1_2})
\cdot (\overline{v^1_1} + \overline{v^1_3})
\cdot (\overline{v^1_1} + \overline{v^1_4})
\cdot (\overline{v^1_1} + \overline{v^1_5})
\cdot (\overline{v^1_2} + \overline{v^1_3})
\cdot (\overline{v^1_2} + \overline{v^1_4})
\cdot (\overline{v^1_2} + \overline{v^1_5})
\cdot (\overline{v^1_3} + \overline{v^1_4})
\cdot (\overline{v^1_3} + \overline{v^1_5})
\cdot (\overline{v^1_4} + \overline{v^1_5})$
Any selection of two or more predictors for gene 1 will result in a 
unsatisfiable solution. Because the $c^1_i$ clause ensures at least one
predictor will be chosen, $c^2_i$ forces our selection to choose
at most one predictor gene $i$. Then constraint $S_2$ 
includes $c^2_i$ so that at most one predictor is chose for each gene.

$S_2 = c^2_1 \cdot c^2_2 \cdot c^2_3$, where

$c^2_1 = 
(\overline{v^1_1} + \overline{v^1_2})
\cdot (\overline{v^1_1} + \overline{v^1_3})
\cdot (\overline{v^1_1} + \overline{v^1_4})
\cdot (\overline{v^1_1} + \overline{v^1_5})
\cdot (\overline{v^1_2} + \overline{v^1_3})
\cdot (\overline{v^1_2} + \overline{v^1_4})
\cdot (\overline{v^1_2} + \overline{v^1_5})
\cdot (\overline{v^1_3} + \overline{v^1_4}) 
\cdot (\overline{v^1_3} + \overline{v^1_5})
\cdot (\overline{v^1_4} + \overline{v^1_5})$

$c^2_2 = 
(\overline{v^2_1} + \overline{v^2_2})
\cdot (\overline{v^2_1} + \overline{v^2_3})
\cdot (\overline{v^2_1} + \overline{v^2_4})
\cdot (\overline{v^2_2} + \overline{v^2_3})
\cdot (\overline{v^2_2} + \overline{v^2_4})
\cdot (\overline{v^2_3} + \overline{v^2_4})$

$c^2_3 = 
(\overline{v^3_1} + \overline{v^3_2})
\cdot (\overline{v^3_1} + \overline{v^3_3})
\cdot (\overline{v^3_1} + \overline{v^3_4})
\cdot (\overline{v^3_1} + \overline{v^3_5})
\cdot (\overline{v^3_2} + \overline{v^3_3})
\cdot (\overline{v^3_2} + \overline{v^3_4})
\cdot (\overline{v^3_2} + \overline{v^3_5})
\cdot (\overline{v^3_3} + \overline{v^3_4}) 
\cdot (\overline{v^3_3} + \overline{v^3_5})
\cdot (\overline{v^3_4} + \overline{v^3_5})$

\item The last constraint $(S_3)$ requires that each
genes must be used as a predictor for at least one other gene
in the satisfying predictor set. 
A gene that is not used in any predictor does perform any regulation function
 and could be removed from the GRN.
$S_3$ ensures that this does not occur.
 To ensure that gene $g_i$ is used in at
least one other predictor, we form clauses $c^3_i$ which includes all 
predictors that use gene $g_i$ as input. To specify that gene $g_i$ must be 
used, we also include a single variable clause $(x_i)$
to $c^3_i$ and add an additional literal $\overline{x_i}$ to 
the other other clauses in $c^3_i$.
The clause $(x_i)$ requires our solution to include gene $g_i$  
and the
$\overline{x_1}$ in the other clauses of 
$c^3_i$ forces at least one other predictor variable
in $c^3_i$ be selected to satisfy the formula.
For example, the $S_3$ clauses for gene $g_1$ are $c^3_1 = (x_1) 
\cdot (\overline{x_1} + v^1_2 + v^1_3 + v^1_5 +
 v^2_1 + v^2_2 + v^2_5 + v^3_1 + v^3_2 + v^3_3 + v^3_5)$.
To satisfy these clauses, $x_1$ and at least
one other predictor variable in $c^3_1$ must be selected.
Again, $S_3$ includes $c^3_i$ for all genes, so:

$S_3 = c^3_1 \cdot c^3_2 \cdot c^3_3$, where

$c^3_1 =
(x_1) \cdot (\overline{x_1} + v^1_2 + v^1_3 + v^1_5 +
 v^2_1 + v^2_2 + v^2_5 + v^3_1 + v^3_2 + v^3_3 + v^3_5)$

$c^3_2 = 
(x_2) \cdot (\overline{x_1} + v^1_2 + v^1_4 + v^1_5 +
 v^2_1 + v^2_3 + v^2_4 + v^3_2 + v^3_4 + v^3_5)$

$c^3_3 = (x_3) \cdot (\overline{x_1} + v^1_1 + v^1_3 + v^1_4 + v^1_5 +
 v^2_2 + v^2_3 + v^2_4 + v^3_3 + v^3_4 + v^3_5)$

Finally we create the SAT formula $S$ as a conjunction of the $S_i$ 
formulas.

$S = S_1 \cdot S_2 \cdot S_3$
\end{enumerate}

{\bf Step 3:}
Constraints together form the CNF $S$ on which the SAT solver performs an 
All-SAT. The cubes
(each cube encodes a candidate predictor set)
 from the All-SAT are collected and the process 
repeats for the remaining attractor cycle orderings. From the results,
we find the most likely predictors based on the frequency of occurrence
of the predictors across
all orderings. Three methods are used to analyze the statistical
results, which will be described in Section ~\ref{sec:expt}.

In general, the algorithm can be applied to input data for $N$ genes and 
$A$ attractor states. 
The total number of attractor
state orderings is $A!$. For each ordering, there can be up to $O(N^3)$ 
predictors per gene. Then SAT search space per ordering is on the 
order of $O(2^{N^3})$ resulting
in overall complexity of $O(A!2^{N^3})$.
Typically, the number of attractor states $A$ recorded through gene expression
measurements are small. 
As such, $A!$ is thus much smaller than $2^{(N^3)}$,
 so the runtime complexity
is dominated by the All-SAT operation. For pragmatic reasons, our
algorithm stops each All-SAT after $T$ minutes (or $C$ cubes), where $T$
or $C$ is 
defined by the user. 

The SAT solver used in our algorithm is based on MiniSat~\cite{minisat}. 
We modify MiniSat
to perform All-SAT optimized for predictor inference with two 
main changes. First, we loop the SAT solving process internal to MiniSat
automatically complementing satisfying assignments
(cubes) and appending the resulting clause to the 
CNF. Second, we modify MiniSat to randomly select branch-variables during
the solving process.
Because MiniSat is originally designed for
finding a single satisfying assignment,
MiniSAT uses a decision heuristic for determining
variables of the final solution. However, this heuristic
will result in many of
the same variables being chosen over iterative runs of MiniSat.
To increase the activity of all variables, we change the random variable
frequency of MiniSat to 100\% (from 2\% in the unaltered MiniSat code) 
to force MiniSAT to always choose a random
variable on every variable-branch decision. A random variable freqeuncy of
$f\%$ means that MiniSat selects the next variable randomly $f\%$ of the
time.

\begin{figure}
\begin{center}
\epsfig{figure=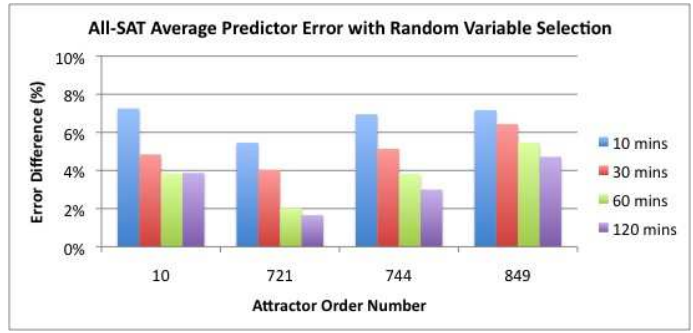, width=0.48\textwidth}
\caption{Average predictor error difference on
melanoma attractor data using MiniSat
with random variable selection modification}
\label{fig:allsat}
\end{center}
\end{figure}

\begin{figure}
\begin{center}
\epsfig{figure=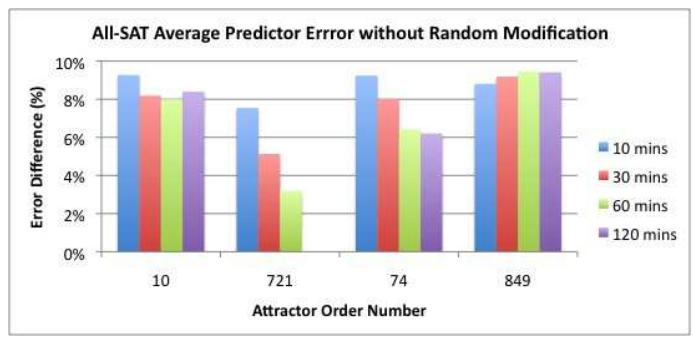, width=0.48\textwidth}
\caption{Average predictor error difference on
melanoma attractor data using MiniSat without modification}
\label{fig:noranallsat}
\end{center}
\end{figure}

To confirm the quality of predictor selection of our modified All-SAT,
 our algorithm was run on 
four selected attractor cycle orderings (labeled 10, 721, 744, and 849)
using melanoma data from~\cite{bittner00},
allowing  the All-SAT operation
to run for 12 hours 
or until all cubes were found, whichever was first.
In the case of attractor cycle order 721, all cubes were found under 12 hours.
We assume that 
12 hours of runtime produce predictor results closely identical to
a complete All-SAT.
In Figure~\ref{fig:allsat} and Figure~\ref{fig:noranallsat},
we compare the average difference in predictors frequency of the 
12 hour (or complete All-SAT) results
with the results obtained with shorter All-SAT runtimes (of 10,
30, 60, and 120 minutes). 
Figure~\ref{fig:allsat} shows the average error difference of all predictors
for the four orderings using MiniSat with the random variable selection 
modification (100\% random variable frequency), while 
Figure~\ref{fig:noranallsat} shows the same MiniSat results without
random variable selection (2\% random variable frequency).
For example, with attractor cycle order 721, 
predictor $f_1=\{g_3,g_5,g_7\}$ had a frequency of occurrence of 
50\% with a 12 hour runtime. Using random variable selection and 
a 30 minute runtime, the same predictor had 43\% occurrence, resulting in a 
difference of 7\%. Without random variable selection, the predictor had
a 69\% occurrence, a 19\% error. 
Across the four orderings analyzed, the average error difference 
over all predictors (shown in Figures~\ref{fig:allsat} and~\ref{fig:noranallsat})
is significantly lower using the
random variable selection modification than without. At 120 minutes,
the random variable selection method has a 
predictor occurrence that differs
from true All-SAT by about 3\%, while without random modification, the difference
is about 8\%.
From this experiment, 
we determine that 30 minutes with random variable selection was sufficient 
to achieve an average of $\leq$ 5\% difference from the true All-SAT results.

\section{Experimental Results}
\label{sec:expt}

\begin{figure*}
\begin{center}
\epsfig{figure=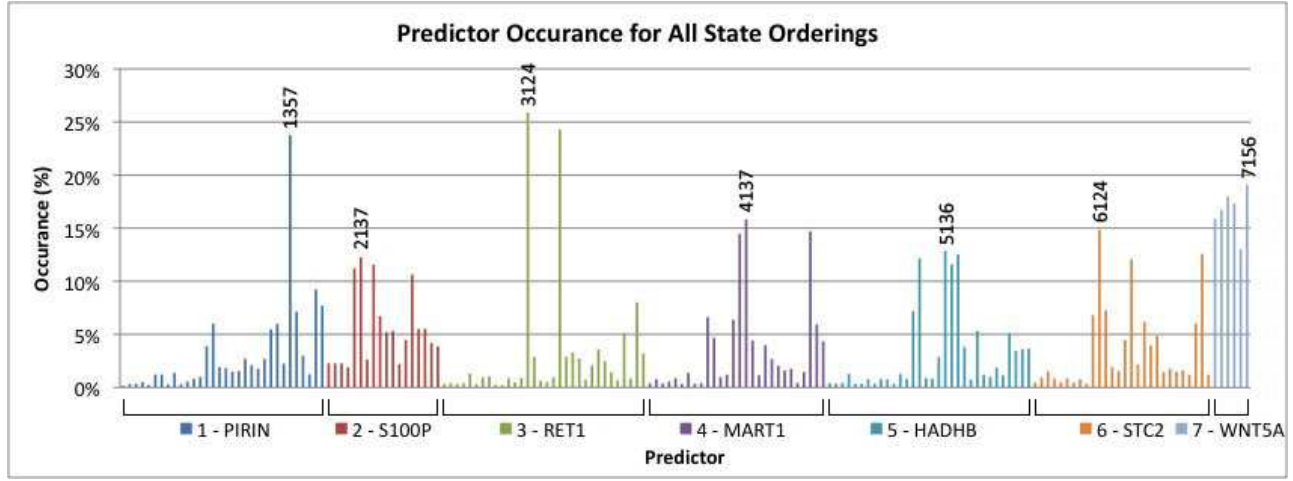, width=0.9\textwidth}
\caption{Method A: Predictor occurrence for all valid attractor cycle orderings
(first iteration: no predictor selected)}
\label{fig:melanoma_hist}
\end{center}
\end{figure*}

To evaluate our SAT-based algorithm for inferring gene predictors, the
algorithm was tested on gene-expression data from a melanoma study done by
Bittner and Weeraratna~\cite{bittner00}. In the melanoma study, it was
observed that an abundance of RNA (expression) for gene WNT5A was
associated with a high metastatis of melanoma. The study measured 587 genes
with 31 gene expression patterns (lines). Seven genes are believed to be
closely knit: PIRIN, S100P, RET1, MART1, HADHB, STC2, and WNT5A. 
There are 18 distinct patterns, which were reduced to seven using
Hamming-distance of one in Table~\ref{tab:melanoma}. 
These seven lines form the attractor states which are the
input to our algorithm.

Our algorithm utilizes a modified open-source and
highly efficient exact SAT-solver called MiniSAT v1.14~\cite{een04, minisat}. 
All-SAT operations were limited to 30 minute time-out. 
On average, each All-SAT run yielded 10K satisifying cubes.
Our algorithm was implemented and run on a Pentium 4 
Linux machine with 4GB RAM.

For the experiments, 
we assume two additional restrictions on attractor cycle orderings.
First, we divide attractor states into good and bad states based on the 
presence of WNT5A. We allow good attractor states to cycle only to other good 
attractor states, and bad attractor states can only cycle to
 other bad attractor states. 
Second, we limit the attractor cycle length to 3 or less since long attractor
cycles are highly complex and unlikely to occur in most biological systems.

In Figure~\ref{fig:melanoma_hist}, we display a histogram of 
all valid predictors and their
frequency of occurrence
over all attractor orderings.
In this chart and table of results, 
a predictor label of 2367 means that gene $g_2$
is predicted by genes $g_3,g_6,$ and $g_7$.
 From this chart, we can observe that
certain predictors occur with significantly higher frequency
than others. For example 
with gene $g_1$,
the predictor $\{x_3,x_5,x_7\}$ 
(PIRIN predicted by RET1, HADHB, WNT5A) occurs with much higher frequency than
all other predictors for gene $g_1$. This indicates that
 this predictor is most likely
to be present in the final predictor set. 

From this data, we propose two methods (A and B) for selecting the predictor
set. 
In {\bf method A}, a predictor histogram is created as in Figure
~\ref{fig:melanoma_hist}. 
From the histogram, for each gene $g_i$, we find its predictor $p^i_j$
such that:

\begin{itemize}
\item $p^i_j$ is the most frequently occurring predictor of gene $g_i$.
\item The {\em resolution ratio} $R_i$ of this predictor (defined as the
ratio of the occurrence frequency of $p^i_j$ to the occurrence 
frequency of the next most frequently ocurring predictor of gene $g_i$)
is maximum.
\end{itemize}

Among all genes, we choose the one with the highest resolution
ratio, and select its most frequently occuring predictor as its
final predictor.
After selecting this final predictor, 
regenerate the histogram, discarding any
 cubes that do not contain the final predictor(s) that have been selected in 
previous steps.
The process repeats until all genes 
have a single final predictor.
The set of final predictors of all genes forms the predictor set. 
The advantage of
method A is that at every iteration, we select real predictors that
have a high overall occurrence in the solution. However the method
may have problems selecting final predictors if
the resolution ratio is low (i.e. when the frequencies
of occurrence of the
predictors are nearly identical).

As an alternative, {\bf method B} is proposed to determine for each gene $i$, 
how likely it is that gene $g_i$ will predict the other genes in the GRN. 
In other words, we ask
what is the occurrence frequency of $x_i$ in the predictors of $f_j$.
Table~\ref{tab:melanoma_tab} 
shows how frequently a gene $g_i$ is
used to predict a gene $g_j$. This table is populated by summing
the occurrence frequency of all predictors of $g_j$ that have gene $g_i$ as one 
if their inputs. 
As such, any entry can be $\geq$100, and is a measure of the usefulness of
$g_i$ as a predictor for $g_j$.
This is done by finding, for each column $j$ of Table~\ref{tab:melanoma_tab},
the three largest entries and adding their values. Suppose we call this value
$s_j$ or the resolution score of column $j$. We compute the resolution 
score for all columns and find the final predictor the colum with the 
highest resolution score. This is done by listing the 3 input genes that
correspond to the 3 entries that were used to compute the highest 
resolution score. 
Similar to method A, we reiterate the process by regenerating the table
after discarding 
all cubes that do not contain predictors that were selected in previous steps.
Method B has the advantage of being more robust
when no {\em single} predictor has a significantly higher 
occurrence frequency than others.
However, there is no guarantee that the predictor selected by method B
is a valid predictor.

\begin{table}[tb]
\scriptsize
\begin{center}
\begin{tabular}{|c|c|c|c|c|c|c|c|} \hline
        & PIRIN & S100P & RET1 & MART1 & HADHB & STC2 & WNT5A \\
        & $x_1$ & $x_2$ & $x_3$ & $x_4$ & $x_5$ & $x_6$ & $x_7$ \\ \hline
BAD     & 0 & 0 & 0 & 0 & 0 & 1 & 1 \\ \cline{2-8}
        & 0 & 0 & 1 & 1 & 1 & 1 & 1 \\ \cline{2-8}
        & 1 & 0 & 1 & 0 & 0 & 0 & 1 \\ \hline
GOOD    & 0 & 1 & 0 & 0 & 0 & 0 & 0 \\ \cline{2-8}
        & 0 & 1 & 1 & 1 & 0 & 0 & 0 \\ \cline{2-8}
        & 1 & 0 & 1 & 1 & 1 & 1 & 0 \\ \cline{2-8}
        & 1 & 1 & 0 & 1 & 1 & 0 & 0 \\ \hline
\end{tabular}
\end{center}
\caption{Attractors for Melanoma Network}
\label{tab:melanoma}
\end{table}

\begin{table}[tb]
\footnotesize
\begin{center}
\begin{tabular}{|c|c|c|c|c|c|c|c|} \hline
        & $f_1$ & $f_2$ & $f_3$ & $f_4$ & $f_5$ & $f_6$ & $f_7$ \\ \hline
$x_1$ &    & 59 & 68 & 57 & 69 & 60 & 19 \\ \hline
$x_2$ & 24 &    & 41 & 29 & 33 & 49 & 51 \\ \hline
$x_3$ & 65 & 48 &    & 76 & 58 & 56 & 17 \\ \hline
$x_4$ & 39 & 40 & 78 &    & 54 & 44 & 29 \\ \hline
$x_5$ & 56 & 30 & 27 & 44 &    & 39 & 54 \\ \hline
$x_6$ & 42 & 54 & 52 & 41 & 44 &    & 86 \\ \hline
$x_7$ & 64 & 63 & 24 & 48 & 32 & 45 &    \\ \hline
\end{tabular}
\end{center}
\caption{Method B: Gene occurrence for all predictors (first iteration:
no predictor selected)}
\label{tab:melanoma_tab}
\end{table}

In our experiments, we also use a hybrid {\bf method AB} which works
in the following manner.
Both methods A and B are used to select their best predictor. If both methods
produce the same predictor $f_i$, we select this predictor as a final
predictor. 
If not, we list the best predictors for each gene for both methods.
If multiple predictors match for both methods, we choose the final
predictor as the one with the highest weighted sum of the resolution
ratio and resolution score. The resolution ratio is weighted by 0.3
and the resolution score is weighted by 0.7. The weighting factor for 
resolution ratio is lower since the resolution ratio values of any gene
are often close to 1. In such a situation, we would like to favor method
B.
If no predictor is produced by the previous step, 
we look at the top five predictors of method A for each gene and
calculate
the weighted sum of their
 resolution ratio and resolution score. The predictor with the
highest weighted sum is selected as the final
predictor. The process is reiterated,
regenerating the histogram and table
at each step, by discarding any cubes that do 
not contain any of the previously selected final predictors. With this combined 
approach, we are able to select predictors with a higher 
degree of confidence and robustness.

We ran our experiments on the  melanoma attractor data 
of~\cite{bittner00}
using methods A, B, and AB.
Results are shown in 
Tables~\ref{tab:methodA},~\ref{tab:methodB}, and~\ref{tab:methodAB} 
respectively. Each table shows what predictor was 
selected at each iteration
and the accompanying resolution ratio (or score). 
Using method A, the predictor set
contains 
\{$f_1=\{3,5,7\}$, $f_2=\{1,3,7\}$, $f_3=\{1,4,6\}$, $f_4=\{3,5,7\}$, 
$f_5=\{1,2,4\}$, $f_6=\{1,2,4\}$, $f_7=\{1,2,4\}$\}.
The predictor set selected by method B contains 
\{$f_1=\{3,5,7\}$, $f_2=\{1,3,7\}$, $f_3=\{1,4,6\}$, $f_4=\{1,3,7\}$,
$f_5=\{1,3,4\}$, $f_6=\{1,3,7\}$, $f_7=\{1,2,6\}$\}.
Finally, the predictor set determined through combining method A and B 
results in
\{$f_1=\{3,5,7\}$, $f_2=\{3,6,7\}$, $f_3=\{1,4,7\}$, $f_4=\{1,3,7\}$,
$f_5=\{1,3,7\}$, $f_6=\{3,5,7\}$, $f_7=\{1,2,4\}$\}. 

\begin{table}[tb]
\scriptsize
\begin{center}
\begin{tabular}{|c|c|c|c|c|c|c|c|} \hline
        & \multicolumn{7}{c|}{Iteration} \\ \cline{2-8}
        & 1 & 2 & 3 & 4 & 5 & 6 & 7 \\ \hline
Predictor selected & 1357 & 6124 & 3146 & 7124 & 5124 & 4357 & 2137 \\ \hline
Resolution ratio   & 2.57 & 1.66 & 1.34 & 1.31 & 1.41 & 1.30 & 1.41 \\ \hline
\end{tabular}
\end{center}
\caption{Predictor set selection using method A}
\label{tab:methodA}
\end{table}

\begin{table}[tb]
\scriptsize
\begin{center}
\begin{tabular}{|c|c|c|c|c|c|c|c|} \hline
        & \multicolumn{7}{c|}{Iteration} \\ \cline{2-8}
        & 1 & 2 & 3 & 4 & 5 & 6 & 7 \\ \hline
Predictor selected & 7126 & 3146 & 5134 & 4137 & 2137 & 1357 & 6137 \\ \hline
Resolution score & 2.56 & 1.84 & 1.99 & 1.97 & 1.77 & 1.78 & 1.98 \\ \hline 
\end{tabular}
\end{center}
\caption{Predictor set selection using method B}
\label{tab:methodB}
\end{table}

From the experiment data, we can draw several conclusive results:
\begin{itemize}
\item It should be noted that the final predictor set from each
method is a valid satisfying
cube of the SAT formula $S$. The iterative steps 
in regenerating the histogram (or table) retain only cubes that contain
previously selected final predictors. 
\item The algorithm enables us to 
generate a few deterministic GRNs. The final predictor
set is present in a select number of attractor cycle orderings. For example,
the final predictor set selected by methods A, B, and AB are found
in respectively 8, 4, and 6 attractor cycle orderings out of the total 
5040 possible orderings.
\item Some predictors are common among
the predictor sets between the three methods. For example, all three methods
select $f_1 = \{g_3,g_5,g_7\}$ (PIRIN predicted by RET1, HADHB, WNT5A).
We can conclude this predictor is highly likely to be a final predictor
in the GRN. 
Also, many that are 
predictors selected by the three methods, while different, 
share common input genes.
For example, 
the exact predictor selected by each method is 
different for gene $g_2$ (S100P), 
but all $f_2$ predictors contain 2 common genes $\{g_3,g_7\}$ (RET1, WNT5A), 
meaning these 2 genes are likely to be contained in the final predictor 
of $f_2$.
\item Using the above results, biologists can target their 
research on gene
regulation and control, focusing on the gene relationships determined
by the predictor set results.
\end{itemize}

\begin{table}[t]
\scriptsize
\begin{center}
\begin{tabular}{|c|c|c|c|c|c|c|c|} \hline
        & \multicolumn{7}{c|}{Iteration} \\ \cline{2-8}
        & 1 & 2 & 3 & 4 & 5 & 6 & 7 \\ \hline
Predictor selected & 1357 & 3146 & 4137 & 7124 & 2367 & 6357 & 5137 \\ \hline
Resolution ratio (A) & 2.57 & 1.07 & 1.11 & 1.57 & 1.28 & 1.77 & 1.01 \\ \hline
Resolution score (B) & 1.85 & 2.04 & 1.83 & 2.01 & 1.70 & 1.23 & 1.69 \\ \hline
\end{tabular}
\end{center}
\caption{Predictor set selection using combined method A and B}
\label{tab:methodAB}
\end{table}

\section{Conclusions}
\label{sec:conclusions}
Determining the predictor set for a gene regulatory network is important
in many applications, particular inference and control of
the GRN.
In this paper, we formulate
gene predictor set inference as an instance of Boolean satisfiability.
In our approach, we determine
all possible orderings of attractor state data,
generate the CNF encapsulating predictor and biological
constraints, and apply a highly-efficient and modified SAT solver
to find candidate predictor sets.
The SAT results are analyzed using three selection methods to produce
the final predictor set. We have tested our algorithm on 
attractor state data from a melanoma study, and determined the 
predictor sets for this GRN.

Encouraged by these results, we plan to expand our SAT-based
algorithm 
to utilize weighted max SAT.
This would provide a more flexible platform where
every predictor has an
associated weight (or importance) in the SAT formulation.
The weighted max SAT algorithm can be tailored for more restrictive
biological constraints,
and also would allow biologists to 
selectively increase or decrease weights on specific predictors. 
This work will incorporate the predictor set results 
to implement an algorithm for the
inference of the complete GRN structure.


\begin{scriptsize}
\bibliography{refs}
\bibliographystyle{ieeetr}
\end{scriptsize}
\end{document}